\documentclass[prd, twocolumn, nofootinbib]{revtex4}
\usepackage{graphicx}
\usepackage{dcolumn}
\usepackage{epsfig} 
\usepackage{amsmath}
\usepackage{amsfonts}
\usepackage{graphicx}
\usepackage{amssymb}
\usepackage{color}

\begin{document}

\title{Improving reconstruction of the baryon acoustic peak : the
  effect of local environment}
\author{I. Achitouv$^{\dagger\ddagger}$\footnote{E-mail:iachitouv@swin.edu.au}}
\author{C. Blake$^{\dagger}$}
\address{$^{\dagger}$Centre for Astrophysics \& Supercomputing,
  Swinburne University of Technology, P.O. Box 218, Hawthorn, VIC
  3122, Australia \\ $^{\ddagger}$ARC Centre of Excellence for All-sky
  Astrophysics (CAASTRO)}

\begin{abstract}

Precise measurements of the baryon acoustic oscillation (BAO) scale as
a standard ruler in the clustering pattern of large-scale structure is
a central goal of current and future galaxy surveys.  The BAO peak may
be sharpened using the technique of density-field reconstruction, in
which the bulk displacements of galaxies are estimated using a
Zel'dovitch approximation. We use numerical simulations to demonstrate
how the accuracy of this approximation depends strongly on local
environment, and how this information may be used to construct an
improved BAO measurement through environmental re-weighting and using
higher-order perturbation theory. We outline further applications of
the displacement field for testing cosmological models.

\end{abstract} 

\maketitle

\section{Introduction}

The large-scale structure of the Universe provides one of our most
powerful tests of the cosmological model, encoding a wealth of
information about the expansion history of the Universe (imprinted as
a standard ruler in baryon acoustic oscillations) and its
gravitational physics (inferred from the growth of structure with
time).  A central goal of modern cosmology is to map out this
structure through galaxy redshift surveys, and to develop accurate
models to link these observations to theory.

In the modelling of cosmological fluid dynamics, a particularly
important role is played by the vector displacement field
$\mathbf{\Psi}(\mathbf{q},t)$, which specifies the trajectory of fluid
elements through space (here denoted by a Lagrangian co-ordinate
$\mathbf{q}$) and time $t$.  In modelling approaches such as
Lagrangian Perturbation Theory, perturbative solutions may be
formulated for $\mathbf{\Psi}$ and used to construct models for the
statistics of the observed density field
(e.g.\citep{Matsubara2008,PadmanabhanWC2009,CarlsonRW2013,Chan2014})

The displacement field has assumed particular importance in
``density-field reconstruction'' (\cite{Eisensteinetal.2007}), a
technique used to sharpen measurements of the baryon acoustic peak in
the galaxy correlation function.  Measurements of the baryon acoustic
peak, a preferred clustering scale imprinted in the distribution of
photons and baryons by the propagation of sound waves in the
relativistic plasma of the early Universe (\citep{PeeblesandYu1970,
  EisensteinandHu1998}), have assumed particular importance in recent
years as a robust standard ruler to map out the cosmic expansion
history (\citep{BlakeandGlazebrook2003,SeoandEisenstein2003}).
However, the pristine sound-horizon scale imprinted in the
high-redshift matter distribution is ``blurred'' by the bulk
displacement of galaxies from their initial positions
(\citep{SeoandEisenstein2007,Seoetal2008,Smitheal2008,Anguloetal2008,CrocceandScoccimarro2008,Sanchezetal2008}).
Reconstruction seeks to use the observed density field to estimate
these displacements, for example using the Zel'dovitch approximation
(\citep{Zeldovitch1970, White2014}) (ZA), and hence retract galaxies
to their near-original positions in the density field, restoring the
sharp preferred separation.  The technique has been successfully
applied to datasets from the Sloan Digital Sky Survey
(\cite{Padmanabhanetal2012}), the Baryon Oscillation Spectroscopic
Survey (\cite{Andersonetal2014}) and the WiggleZ Dark Energy Survey
(\cite{Kazinetal2014}).

A series of studies has considered the theory and implementation of
density-field reconstruction.  It may be formulated within Lagrangian
Perturbation Theory
(\citep{PadmanabhanWC2009,NohWhitePadmanabhan2009}), to which the ZA
is the lowest-order contribution, or by alternative perturbative
schemes (\cite{McCullaghSzalay2012,TassevZaldarriaga2012}), or in
Fourier space (\cite{BurdenPercivalHowlett2015}).  The method is
robust to the treatment of the bias of the tracer or redshift-space
effects
(\citep{Seoal2010,Mehtaetal2011,Padmanabhanetal2012,Burdenetal2014,VargasMaganaetal2014}),
although room for improvement in the algorithm certainly exists
(\cite{TassevZaldarriaga2012,White2015}).

In this paper we undertake a forensic study of the performance of the
current reconstruction algorithm through comparison with the exact
displacement field derived from particle-tracking in N-body
simulations.  First, we study the accuracy of the ZA using the initial
matter velocity field with different hypotheses for the smoothing
procedure and its sensitivity to the local density. This fiducial
investigation is not affected by complications due to galaxy bias and
non-linear effects on the estimated density field.  Using this to
create an unbiased understanding of how proto-halo displacement is
optimally modelled, we extend our results to the reconstruction of the
displacement from the non-linear low-redshift halo density field.  In
particular, we study the error in the estimated displacements as a
function of galaxy halo mass, redshift, filter and smoothing scale for
estimating the density field.  We also also study how this error
depends on the order of perturbation theory applied, and local
environment.  We discuss the consequences of these results for the
optimal measurement of the distance scale and the use of the
displacement field to test cosmological models.  For the purposes of
this investigation, we neglect redshift-space distortions in the
galaxy co-ordinates.

Our paper is structured as follows: in Section II we review the
connection between the displacement and density fields within
Lagrangian Perturbation Theory. We describe our methods for recovering
the exact displacement field from N-body simulations, and the
estimated displacement field from the initial simulation velocity
field.  In Section III we examine the performance of the density-field
reconstruction algorithm, in particular focussing on the degradation
of this performance with increasing local density.  In Section IV we
suggest how the measured post-reconstruction correlation function can
be ``re-weighted'' by environment to maximize the sharpness of the
baryon acoustic peak, and we consider the level of improvement which
may be obtained in the fitted distance scale.  In Section V we discuss
the implications of our findings for the use of density-field
reconstruction, and the displacement field itself, for testing
cosmological models.

\section{Displacement Field: theory and measurement}
\label{sec2}

\subsection{The Zel'dovitch and 2LPT approximations}

In this section we briefly review Lagrangian Perturbation Theory (LPT)
at first (i.e., ZA) and second order (i.e., 2LPT). The displacement
vector field $\mathbf{\Psi}$ links an initial (Lagrangian) position
$\mathbf{q}$ of a mass element $M$, smoothed on a scale $R_S(M)$, to
the current (Eulerian) position,
\begin{equation}
\mathbf{x}(R_S)=\mathbf{q}(R_S)+\mathbf{\Psi}(\mathbf{q},R_S).
\end{equation}
In the ZA, $\mathbf{\Psi}$ is determined by the gradient of the
gravitational potential $\nabla \Phi$, which can be expressed in terms
of the linearly extrapolated density field.  Considering only the
scalar contribution $\mathbf{\Psi}$, at first and second order in the
matter density field, we have (\citep{Chan2014})
\begin{equation}
\begin{split}
&\mathbf{\Psi} \simeq \mathbf{\Psi}^{(1)} + \mathbf{\Psi}^{(2)} \\
&\mathbf{\Psi} \simeq -D \nabla \phi^{(1)} - \frac{3}{7} D^2 \nabla \phi^{(2)}
\end{split}
\label{2LPTdisp}
\end{equation}
where $D$ is the linear growth factor and the gravitational potential
can be solved using the following Poisson equations:
\begin{equation}
\begin{split}
&\nabla^2 \phi^{(1)}(R_S) = \delta_0(R_S) \\
&\nabla^2 \phi^{(2)}(R_S) = -\frac{1}{2} \mathcal{G}_2(\phi^{(1)})
\label{Poisson}
\end{split}
\end{equation}
where $\delta_0(R_S)$ is the linear density field smoothed on scale
$R_S$ and
\begin{equation}
\mathcal{G}_2(\phi^{(1)})=\sum_{i,j} \left[ (\nabla_{ij}
  \phi^{(1)})^2 - (\nabla^2 \phi^{(1)})^2 \right] .
\end{equation}
The first order displacement ($^{(1)} $ term) is the well-known
ZA which can be expressed as a function of the
initial peculiar velocity
\begin{equation}
\mathbf{v}(R_S) = a \partial_t D(z) \partial_D \mathbf{\Psi}^{(1)} ,
\label{velZA}
\end{equation}
where $a$ is the cosmic scale factor.  The second order term has the
label $^{(2)}$. The variance of the displacement field in the ZA is
directly proportional to the initial matter power spectrum
\citep{Chan2014} such that (e.g.\ for the $x$-component)
\begin{equation}
\langle \Psi_{x}^{(1)}(R_S)^2 \rangle \propto D^2 \int
\tilde{W}^2(k,R_S) \, P_L(k,z_i) \, dk ,
\label{varpsiZA}
\end{equation}
where $P_L$ is the linear matter power spectrum at initial redshift
$z_i$ and $\tilde{W}$ is the Fourier transform of the smoothing
function.  The higher orders of the displacement field can be
expressed as higher-dimensional integrals \citep{Chan2014}.

In order to determine the accuracy of the ZA in various scenarios, we
test two cases:

\begin{itemize}

\item Case 1: we use the proto-halo velocities to predict halo
  positions at $z=0$ assuming the ZA, and compare to the exact
  simulation displacement field (see Sec II B,C).

\item Case 2: we use the halo density field at $z=0$, and second order
  LPT, to reconstruct the displacement $-\mathbf{\Psi}$ of halos, and
  compare to the exact simulation displacement field (see Sec III).

\end{itemize}
 
The first case extends the work of \cite{Chan2014} (for the unsmoothed
dark matter density field) to the displacement of proto-halos. In
particular, this analysis provides a physical insight of how
proto-halos are displaced from their positions in the initial matter
density field through cosmic time up to $z=0$.  We also extend these
studies to investigate the sensitivity of the result to the smoothing
scale and local environment, yielding useful comparisons with case 2.

The second case has a powerful application in the reconstruction of
the baryon acoustic peak.  Previous studies have applied this method
using first order LPT and found a low sensitivity to the smoothing
scale $R_S$ which enters into Eq.(\ref{2LPTdisp},\ref{Poisson})
(\citep{Padmanabhanetal2012, Burdenetal2014}).  We extend these tests
to consider the effect of local environment and evaluate the gain of
adding the 2LPT correction to reconstruct the displacement of
halos. In particular, we will show how the second order correction
depends on the choice of smoothing scale and environment.

\subsection{Exact displacement field from N-body simulations}

In order to test the accuracy of the displacement field in cases 1 and
2, we use the DEUS simulations. These simulations were run for several
scientific purposes described in
\cite{Ali,Courtin,Yann,Blot,Raseraetal,AWRW}. The box side of the
simulations is $648 \, h^{-1}$ Mpc and they contain $1024^3$ particles
of mass $\sim 1.75 \times 10^{10} h^{-1} M_{\odot}$. The simulations
were carried out using the RAMSES code \cite{Teyssier2002} for a
$\Lambda$CDM model calibrated to the WMAP 5-year cosmological
parameters \cite{wmap5}.  The halos were identified using the
Friend-of-Friends (FoF) algorithm with linking length $b=0.2$.

In what follows we will use the Lagrangian size of halos $R_L$ where
their mass is given by $M(R_L) = V(R_L) \bar{\rho}$, where
$\bar{\rho}$ is the mean matter density.  We will approximate the
shape of proto-halos to be spherical, with volume $V(R_L) =
\frac{4}{3} \pi R_L^3$.
 
We measured the position of the center of mass $\mathbf{x}_f$ of all
halos at $z=0$.  In order to measure the position of the proto-halos
in the initial conditions, we labelled all particles which belong to a
halo at $z=0$ and compute the center of mass $\mathbf{x}_i$ for these
particles in the initial conditions of the simulation. Therefore we
measured for each halo the exact displacement $\mathbf{\Psi} =
\mathbf{x_f} - \mathbf{x_i}$.

\subsection{Displacement prediction from initial velocity field}

The final position of a proto-halo at $z=0$ may be predicted from the
initial velocity field according to Eq.(\ref{velZA}).  Hence, the
displacement of a mass element contained in a smoothing radius $R_S$
can be expressed as function of the average peculiar velocity
initially contained in a patch of size $R_S$, $\mathbf{v}_i(q,R_S)$
as:
\begin{equation}
\mathbf{\Psi}^{(1)}(q,z,R_S)=\frac{\mathbf{v}_i(R_S) D(z)}{a_i
  H(a_i)f(a_i)D(z_i)}
\label{Psi_Zv}
\end{equation}
where $H(a_i)$ is the Hubble parameter evaluated at the initial scale
factor $a_i$ and $f\equiv d\ln{D}/d\ln{a}$ is the linear growth
rate. In this approximation we can use Eq.(\ref{Psi_Zv}) to displace
the proto-halos by measuring the average velocity $\mathbf{v}_i$
around the center of mass $\mathbf{q}=\mathbf{x}_i$ and within a
smoothing radius $R_S$. We investigated the accuracy of the ZA in
predicting the magnitude of $\Psi \equiv \mid\mathbf\Psi\mid$.

The smoothing scale $R_S$ is a free parameter in our investigation,
used in the estimation of the local overdensity field.  For $R_S
\rightarrow \infty$ the average velocity within the smoothed region
tends to zero, and hence the proto-halos are not displaced.  On the
contrary, as $R_S \rightarrow 0$ we obtain the displacement of
individual particles, without considering that they belong to a halo.

\begin{figure}
\begin{center}
\includegraphics[scale=0.4]{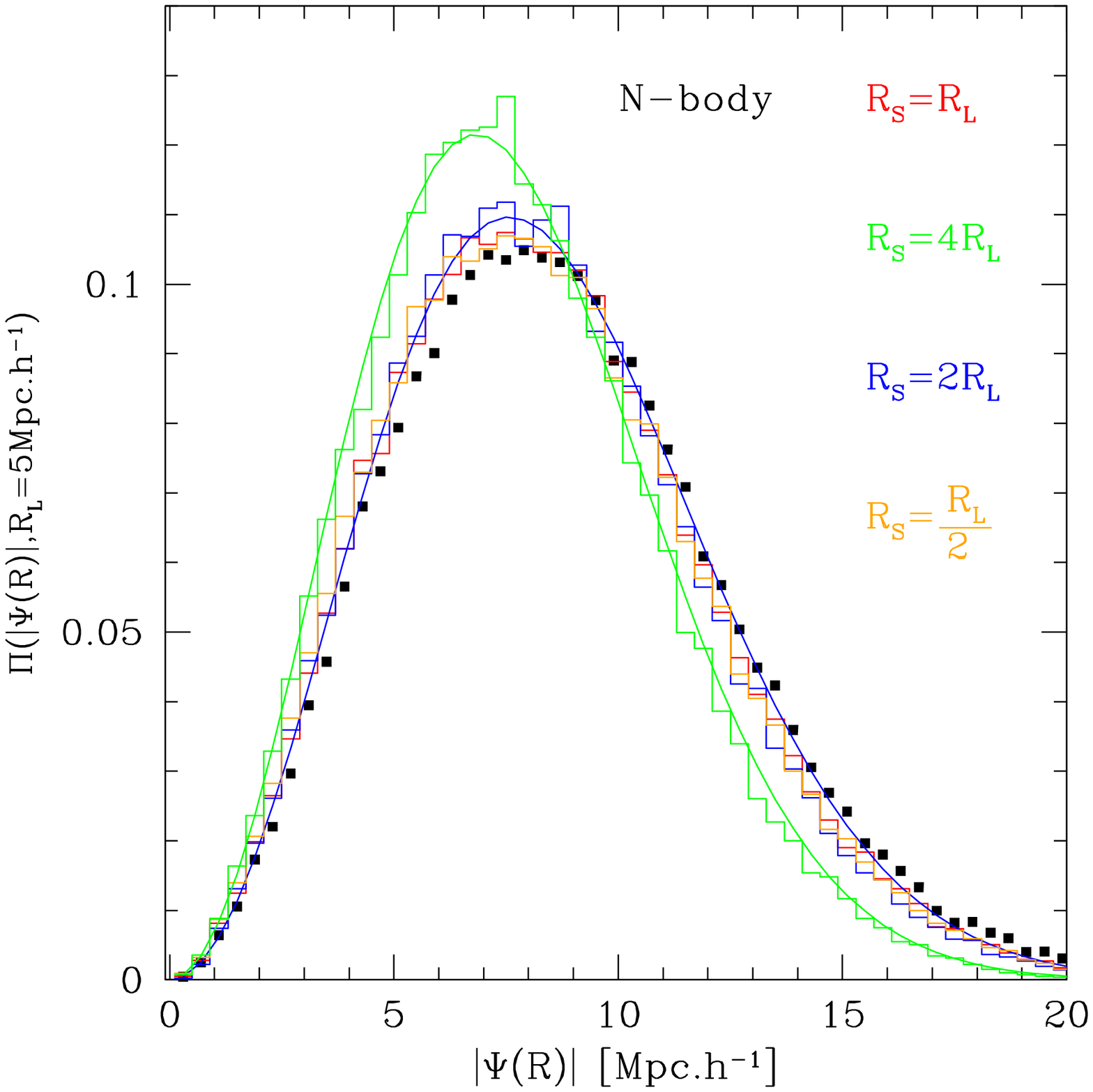}\\
\includegraphics[scale=0.4]{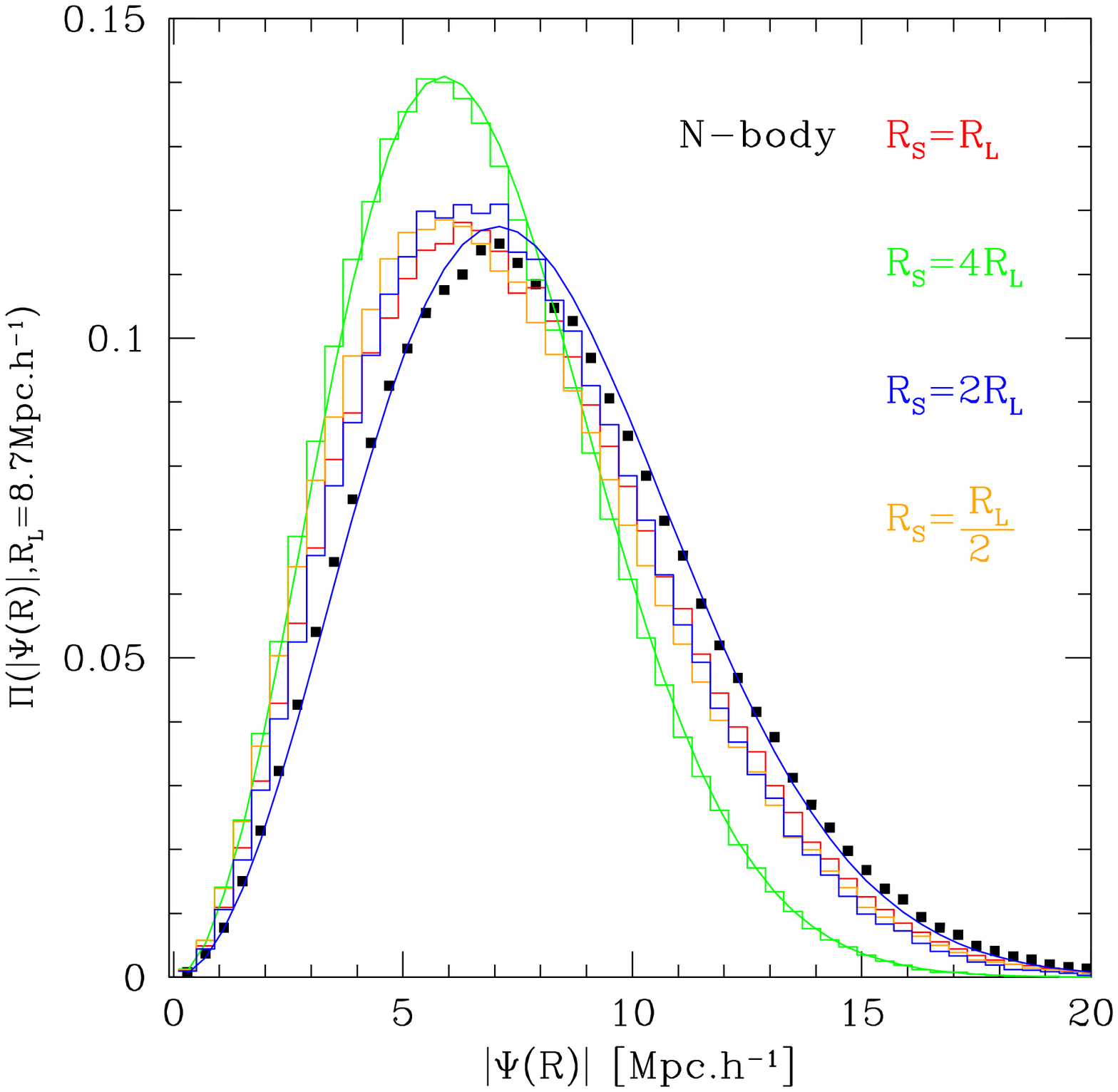}
\caption{The distribution of $\Psi$ measured in N-body simulations for
  two different halo masses (black squares).  The upper panel shows
  halos with Lagrangian radius $R_L = 5 \, h^{-1}$ Mpc while the
  bottom panel shows halos with Lagrangian radius $R_L = 8.7 \,
  h^{-1}$ Mpc.  For each panel, the histograms show the Zel'dovitch
  prediction Eq.(\ref{Psi_Zv}) using different smoothing scales $R_S$
  (see legend). The solid lines correspond to the Maxwell-Boltzmann
  distribution with variance $\sigma_\Psi^2(R_S)$ computed according
  to Eq.(\ref{varZel}).}
\label{Fig1}
\end{center}
\end{figure}

In Fig.(\ref{Fig1}), we show the measured PDF of $\Psi$ (black
squares) for two different masses of halos: the upper panel
corresponds to halos with Lagrangian radii $R_L = 5 \, h^{-1}$ Mpc ($M
= 10^{13.6} h^{-1} M_\odot$) while the lower panel corresponds to
halos with Lagrangian radius $R_L = 8.7 \, h^{-1}$ Mpc ($M = 10^{14.3}
h^{-1} M_\odot$). As we can see, the mean displacement is not strongly
sensitive to the mass of the proto-halos (varying by only $\sim 15
\%$). However, the variance of the absolute displacement is reduced
for large halos.  Physically, this is because larger proto-halos are
less sensitive to the external shear field which contributes to the
overall dynamics. In Fig.(\ref{Fig1}), we can also see the prediction
of Eq.(\ref{Psi_Zv}) (plotted as histograms) using different smoothing
scales $R_S$ (see legend) when measuring the average peculiar
velocities.  For $R_S \leq 2R_L$, the absolute displacement predicted
by Eq.(\ref{Psi_Zv}) is approximately independent of $R_S$.  For $R_S
= 4R_L$, we begin to smooth out relevant density fluctuations and
hence underestimate the absolute displacement.  In the limit $R_S >>
R_L$, the analytical pdf of $\Psi$ is given by a Maxwell-Boltzmann
distribution (\cite{Koppetal2015}),
\begin{equation}
\Pi(\Psi,\sigma_{\Psi})=\sqrt{\dfrac{2}{\pi}}\left(\frac{\sqrt{3}}{\sigma_{\Psi}}
\right)^3 \Psi^2\exp{\left[ -\frac{3\Psi^2}{2\sigma^{2}_{\Psi}}\right]
}
\label{psiabs}
\end{equation} 
where the variance is
\begin{equation}
\sigma^{2}_{\Psi}(R) = \frac{1}{2\pi^2} \int \tilde{W}^2(R,k) \, P_L(k,z) \, dk .
\label{varZel}
\end{equation}
The Fourier transform of the filter function $\tilde{W}(R,k)$ is
chosen to be the Fourier transform of a top-hat filter in real space.
In Fig.(\ref{Fig1}) we can see this analytical prediction for $R=4R_L$
and $R=2R_L$ (solid lines).  For $R=4R_L$ the analytic prediction of
Eq.(\ref{psiabs}) matches the Zel'dovitch prediction of
Eq.(\ref{Psi_Zv}) with $R_S=4R_L$ (comparing the green solid line and
green histogram).  For smaller smoothing scales ($R=2R_L$), the regime
$R>>R_L$ is no longer satisfied and we start observing deviations
between Eq.(\ref{psiabs}) and Eq.(\ref{Psi_Zv}) (comparing the blue
solid line and blue histogram). Interestingly, we observe that
Eq.(\ref{psiabs}) qualitatively follows the trend for the absolute
displacement of proto-halos (black squares) once we use $R=2R_L$. This
observation could be used to model the clustering of halos with high
accuracy, although this goes beyond the scope of this paper.

\begin{figure}
\begin{center}
\includegraphics[scale=0.4]{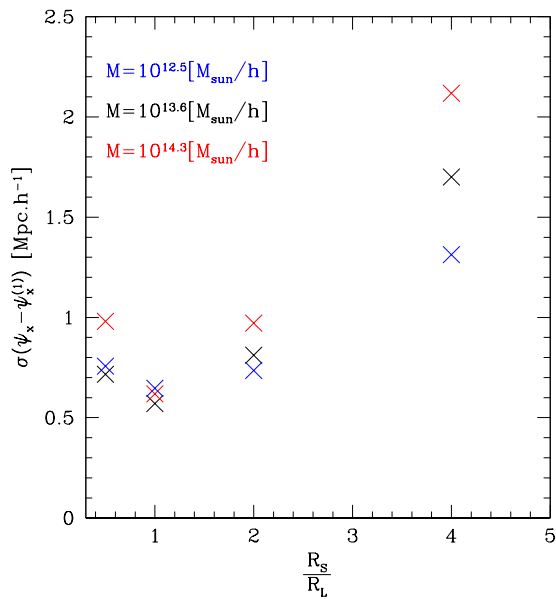}
\caption{Standard deviation of the difference between the
  $x$-component of the exact displacement and the ZA
  Eq.(\ref{Psi_Zv}), as a function of the ratio of the smoothing
  length and Lagrangian size of the halos, $R_S/R_L$.  The red, black
  and blue crosses correspond to halo masses $10^{12.5}$, $10^{13.0}$
  and $10^{14.3} h^{-1} M_\odot$.}
\label{Fig3}
\end{center}
\end{figure}

Second, we consider the dispersion between the exact and estimated
displacements.  For each component $i$, the PDF of $\Psi_i -
\Psi^{(1)}_{i}$ follows a Gaussian distribution with a mean value
equal to zero. In Fig.\ref{Fig3} we can see the standard deviation of
the difference between the exact and the Zel'dovitch prediction
(Eq.(\ref{Psi_Zv})) of $\Psi_x$, for halo masses $10^{12.5}$,
$10^{13.0}$ and $10^{14.3} h^{-1} M_\odot$, as a function of the
smoothing length $R_S$ (similar results are obtained for $\Psi_y$ and
$\Psi_z$).  Independently of the proto-halo masses, the optimal
smoothing scale to predict the final position of halos is $R_S=R_L$.
At this scale $\Psi$ can be determined with an accuracy of $\sim 0.5
\, h^{-1}$ Mpc, a few per cent of the mean displacement.  This is an
impressive result considering that we only use the ZA.  It implies
that if we can identify proto-halos in the initial matter density
field, we can predict their clustering at $z=0$ with high accuracy in
a very short time.

\begin{figure}
\begin{center}
\includegraphics[scale=0.4]{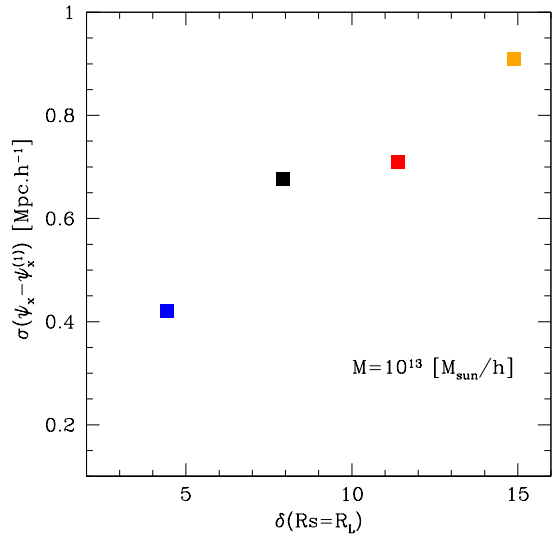}
\caption{Standard deviation of the difference between the exact
  displacement and that predicted by Eq.(\ref{2LPTdisp},\ref{Poisson})
  with $R_S=R_L$ in different environments.  Colors represent
  different environments.  The black square corresponds to the average
  density around halos $\bar{\delta}(R_S=R_L)$.  The blue, red and
  orange squares correspond to environments with local overdensity
  $\delta = \lbrace \bar{\delta}-\sigma_{\delta},
  \bar{\delta}+\sigma_{\delta}, \bar{\delta}+2\sigma_{\delta} \rbrace$
  respectively, where $\bar{\delta}$ is the mean overdensity and
  $\sigma_{\delta}$ is the standard deviation of the overdensity
  distribution amongst the halos.}
\label{Fig4bis}
\end{center}
\end{figure}

\begin{figure}
\begin{center}
\includegraphics[scale=0.4]{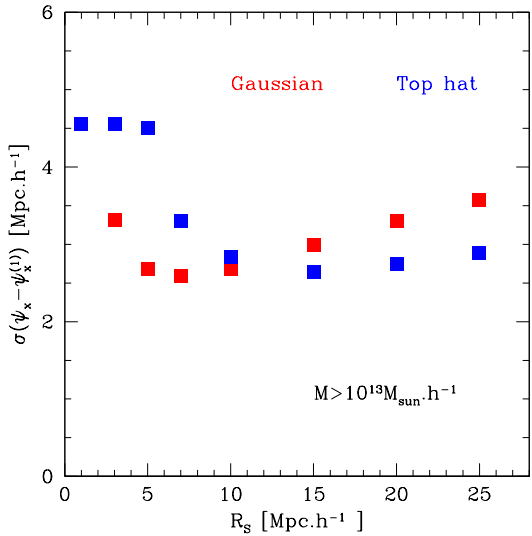}
\caption{Standard deviation of the difference between the
  $x$-component of the exact displacement, and the ZA computed using
  the $z=0$ density field using Eq.(\ref{2LPTdisp},\ref{Poisson}), as
  a function of the smoothing length $R_S$ and the type of filter
  used.}
\label{Fig4}
\end{center}
\end{figure}

Finally, we test how sensitive the ZA (Eq.(\ref{Psi_Zv})) is to the
local environment.  Each environment is defined by the number density
of halos inside a sphere of radius $R_S$.  We divide the halos into
environment bins by considering the probability density function for
the number of halos at $z=0$ within $R_S=R_L$, in particular the mean
$\bar{\delta}$ and standard deviation $\sigma_{\delta}$ of this
distribution.  We then define 4 different environments corresponding
to halos with surrounding densities $\delta = \lbrace
\bar{\delta}-\sigma_{\delta}, \bar{\delta},
\bar{\delta}+\sigma_{\delta}, \bar{\delta}+2\sigma_{\delta} \rbrace$.

In Fig.(\ref{Fig4bis}) we can see the standard deviation between the
true value and the ZA prediction of $\Psi_x$ (with $R_S=R_L$), as
function of the surrounding number density of halos ($\delta(R_S)$),
considering halos with mass $M = 10^{13} \, h^{-1} M_\odot$, which is
equivalent to $R_L \sim 3 \, h^{-1}$ Mpc. The black square corresponds
to the mean overdensity environment.  As we can see, the accuracy of
the ZA prediction is sensitive to the environment.  For overdense
environments, the ZA prediction becomes less accurate compared to
underdense environments.  This suggests that the displacement field of
proto-halos contains non-linear information that can not be
encapsulated in the simple ZA.  An extension of this work would be to
investigate if the sensitivity to the smoothing scale and environment
depends on cosmology.

In the next section, we will repeat this exercise, comparing the
displacements reconstructed using the halo distribution at $z=0$ to
evaluate the density contrast in Eq.(\ref{Poisson}), to the exact
displacements.

\section{Performance of density-field reconstruction}
\label{sec3}

\subsection{Displacement prediction from density-field reconstruction}

We first describe how we predict the displacement field from the $z=0$
halo distribution using the technique of density-field reconstruction.
Dividing the volume into a 3D grid, we first estimated the halo
overdensity distribution at each grid point, using a particular filter
(Gaussian or top-hat) and smoothing scale.  We then convert the halo
overdensity $\delta_h(R_S)$ to a matter overdensity $\delta_m(R_S) =
\delta_h(R_S)/b$ assuming a linear bias factor $b$, which we determine
by comparing the amplitude of the large-scale halo power spectrum to
the linear-theory power spectrum used to generate the simulation.
This allows us to solve Eq.(\ref{Poisson}) using Fast Fourier
Transform methods, and hence determine the displacement field
$-\mathbf{\Psi}(R_S)$ via Eq.(\ref{2LPTdisp}).

\subsection{Dependence on smoothing scale and filter}
\begin{figure}
\begin{center}
\includegraphics[scale=0.4]{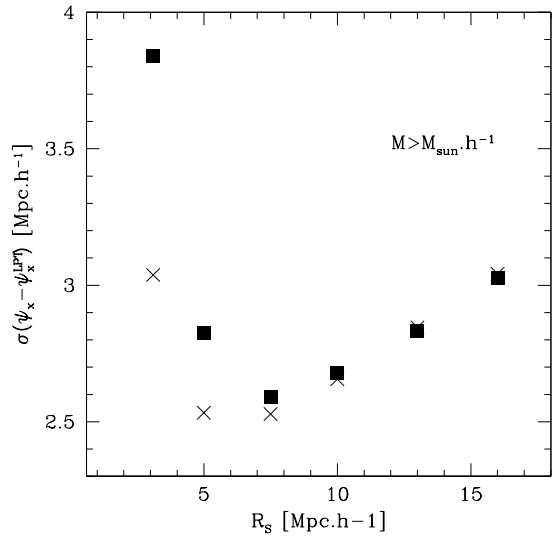}
\caption{Standard deviation of the difference between the
  $x$-component of the exact displacement, and that computed using the
  $z=0$ density field using Eq.(\ref{2LPTdisp},\ref{Poisson}), as a
  function of the smoothing length $R_S$ using a Gaussian
  filter. Black squares show the first-order (ZA) prediction while
  black crosses show the prediction including the 2LPT correction.}
\label{Fig2LPTall}
\end{center}
\end{figure}

Considering only the first order of $\mathbf{\Psi}$ in
Eq.(\ref{2LPTdisp},\ref{Poisson}), we used several smoothing scales in
the range $0 < R_S< 25 \, h^{-1}$ Mpc for halos with a mass $M>10^{13}
h^{-1} M_{\odot}$, typical of Luminous Red Galaxies probed by
large-scale structure surveys such as the Baryon Oscillation
Spectroscopic Survey (\cite{Andersonetal2014}).  We used both a
top-hat filter and a Gaussian filter, and we measured the error made
in the Zel'dovitch prediction compared to the exact values for each
component as a standard deviation $\sigma(\psi_i^{(1)} - \psi_i)$.
The result for the $x$-axis is shown in Fig.(\ref{Fig4}), other axes
produced similar results.  We find that there is again an optimal
smoothing scale which depends on the filter function. The optimal
performance of both filters leads to a residual error $\sigma \sim 2.5
\, h^{-1}$ Mpc, around 5 times poorer than the determination based on
the initial proto-velocities.

For small smoothing scales the performance of the reconstruction
method becomes worse, since the number of neighbours becomes too small
to reconstruct the density field with sufficient accuracy.
Furthermore, the higher order correction in Eq.(\ref{2LPTdisp})
becomes non-negligible when $R_S \rightarrow 0$.  For large smoothing
scales, the degradation from the optimal performance is a slow
function of $R_S$, consistent with the displacement being generated by
large-wavelength modes that may be successfully recovered even in the
presence of significant smoothing (\cite{Eisensteinetal.2007}).  These
effects are illustrated in Fig.(\ref{Fig2LPTall}). Using a Gaussian
filter we show the error made in the 1st-order (black squares) and
2LPT (black crosses) predictions for different smoothing scales.  The
2nd-order correction leads to a $\sim 2 \%$ improvement 
for $R_S \sim 8 h^{-1}$ Mpc and a $\sim 12 \%$ improvement
for $R_S \sim 5 h^{-1}$  Mpc.  
In the next section we will emphasize the connection
between this result and the local environment.

Overall, the choice of filter (Gaussian vs. top-hat) does not affect
the performance of the displacement reconstruction at first-order LPT,
once we choose the optimal smoothing scale (see
Fig.(\ref{Fig4})). Therefore in what follows we will restrict our
analysis to a Gaussian filter.

%

\subsection{Dependence on local environment}
\label{secenv}

\begin{figure}
\begin{center}
\includegraphics[scale=0.4]{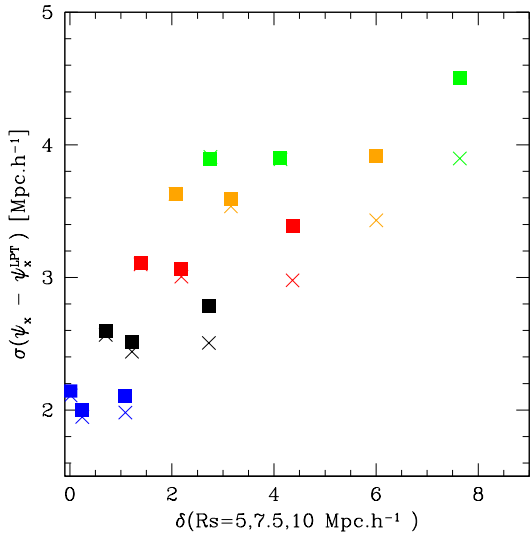}
\caption{Standard deviation of the difference between the exact
  displacement and that predicted by Eq.(\ref{2LPTdisp},\ref{Poisson})
  based on the $z=0$ halo density field.  The squares show the first
  order implementation of Eq.(\ref{2LPTdisp},\ref{Poisson}) while the
  crosses implement the second order correction.  Colors represent
  different environments defined in the text (from bottom to top: E1,
  E2, E3, E4 and E5). For each color we choose 3 smoothing scales
  (from left to right: $R_S = 10, 7.5, 5 \, h^{-1}$ Mpc.}
\label{Fig5}
\end{center}
\end{figure}

In this Section we investigate how the accuracy of the reconstructed
displacement field depends on the local environment of each halo.
Given a smoothing scale $R_S$, we can compute the number of neighbours
for each halo and therefore estimate the local density contrast.  We
used the mean local density around halos $\bar{\delta}$, and the
standard deviation $\sigma_\delta$, to define five bins in environment
E1, E2, E3, E4 and E5 with central values $\delta = \left\lbrace
\bar{\delta}-\sigma_\delta, \bar{\delta}, \bar{\delta}+\sigma_\delta,
\bar{\delta}+2\sigma_\delta, \bar{\delta}+3\sigma_\delta
\right\rbrace$.  For each local environment we measured the error in
the reconstructed displacement at different smoothing scales.

In Fig.(\ref{Fig5}) we show the standard deviation of the difference
between the reconstructed and exact $\Psi_x$, as function of the
surrounding smoothed halo density $\delta(R_S)$.  The blue, black,
red, orange and green data points correspond respectively to
environments E1, E2, E3, E4, and E5. For each environment we plot 3
different squares and 3 different crosses which correspond to three
different smoothing scales $R_S = 10, 7.5, 5 \, h^{-1}$ Mpc, from left
to right. The squares correspond to the prediction of first-order LPT,
while the crosses implement the second-order correction.

In agreement with Fig.(\ref{Fig4bis}), the accuracy of the prediction
is highly sensitive to the environment. Furthermore, in
Fig.(\ref{Fig5}) we can see the effect of the second-order
correction. Unsurprisingly, if we choose a large smoothing scale
(e.g.\ $10 \, h^{-1}$ Mpc), the second order correction in
Eq.(\ref{2LPTdisp}) is negligible.  For a smaller smoothing scale
(e.g.\ $R_S = 5 \, h^{-1}$ Mpc), the second-order correction becomes
more important.  In fact, the second-order correction contains a term
proportional to $\delta^2(R_S)$ and for small values of $R_S$ the
average local density around halos increases
($\bar{\delta}(R_S=10)\sim 0.8$, $\bar{\delta}(R_S=5)\sim 2.8$).
However we can observe that even for $R_S = 5 \, h^{-1}$ Mpc, the
improvement in the residual from the second-order correction is less
significant than the difference in residual between E1 and E2.  This
suggests that the displacement contains non-linear information that
can not be described by deterministic corrections.

Finally, for dense environments (e.g $\delta >\bar{\delta}$), a
smaller smoothing scale leads to a better reconstruction of the final
halo position predicted by the 2LPT approximation
(Eq.\ref{Poisson}). For halos in a region with the mean density or
below, reducing the smoothing scale does not particularly improve the
accuracy of the displacement field. This convergence of the optimal
smoothing scale toward $R_S \rightarrow R_L$ shows the limits of the
reconstruction method.  Reconstruction applied at $z=0$ on the
non-linear halo density field is not equivalent to predicting the
final position of halos from the initial conditions.

\subsection{Dependence on redshift and halo mass}

\begin{figure}
\begin{center}
\includegraphics[scale=0.4]{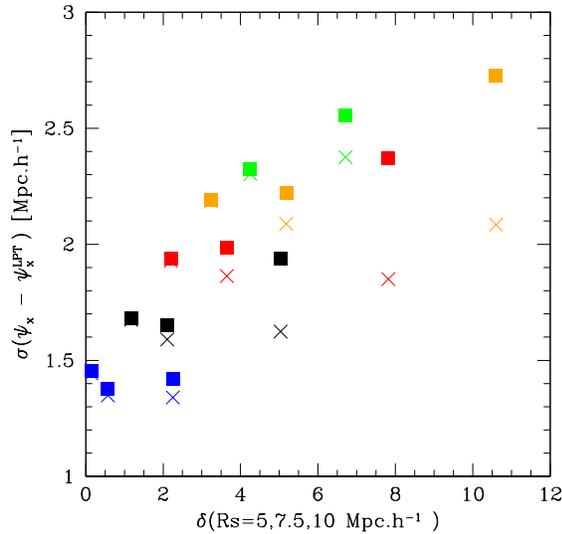}
\caption{Same as Fig.(\ref{Fig5}), for halos at redshift $z=1$ in the
  mass range $M > 10^{13} h^{-1} M_\odot$.}
\label{Fig6}
\end{center}
\end{figure}

We repeated the previous analysis at redshift $z=1$.  Fig.(\ref{Fig6})
displays these results in the same format as Fig.(\ref{Fig5}).  As
before, for large values of the smoothing (e.g. $R_S = 10 \, h^{-1}$
Mpc), the second-order correction is negligible, while it becomes
important for smaller smoothing scales when the local density around
halos is high. In this case, choosing a smaller smoothing scale
(e.g. $R_S = 5 \, h^{-1}$ Mpc) and adding the second-order correction
to Eq.(\ref{Poisson}) gives a better description of the displacement
field (Eq.(\ref{2LPTdisp})).  We note that at $z=1$ the second-order
correction is more important than at $z=0$. This is due to the choice
of halos we consider. At $z=1$, halos with $M > 10^{13} h^{-1}
M_\odot$ are more biased and belong to exponential tail of the halo
mass function \citep{AWWR}.  Hence the local overdensity around those
halos is larger, leading to a higher second-order correction of the
displacement field.

\begin{figure}
\begin{center}
\includegraphics[scale=0.4]{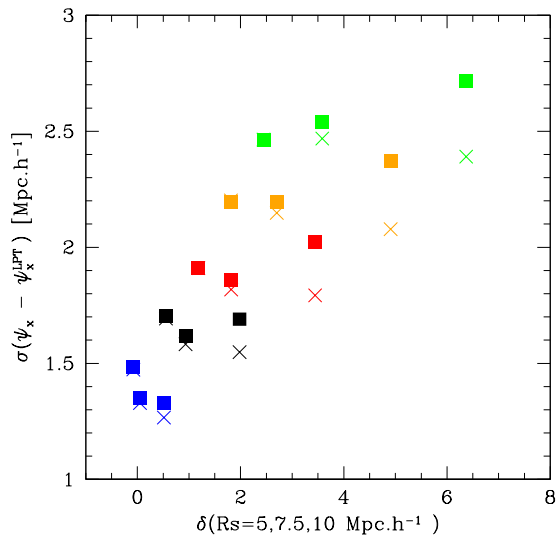}
\caption{Same as Fig.(\ref{Fig5}), for halos at redshift $z=1$ in the
  mass range $M > 10^{12.5} h^{-1} M_\odot$.}
\label{Fig6bis}
\end{center}
\end{figure}

For halos with mass $M > 10^{12.5} h^{-1} M_\odot$ at $z=1$, the error
in the displacement is shown in Fig.(\ref{Fig6bis}).  For $R_S = 10 \,
h^{-1}$ Mpc, the second-order correction is negligible and the error
in the displacement is similar to that found for halos with $M >
10^{13} h^{-1} M_\odot$ at $z=0$.  For smaller smoothing scales, the
second-order correction becomes more important.

Overall, at $z=1$ the analytic approximations for the displacements
(Eq.(\ref{2LPTdisp})) are in better agreement with the true
displacements we measure in the simulations, by a factor which depends
on the environment. For instance at the average density (E3, black
cross), the standard deviation of the residual is $\sim 2.5 \, h^{-1}$
Mpc at $z=0$ and $\sim 1.6 \, h^{-1}$ Mpc at $z=1$.

\section{Effect on the baryon acoustic correlation function peak}
\label{sec4}

In this Section we study how the reconstruction of the baryon acoustic
peak depends on the accuracy with which the displacement field can be
determined. In particular, we investigate if the ``sharpness'' of the
reconstructed acoustic peak depends on local overdensity, and if this
effect may be used to obtain improved accuracy in the resulting
standard ruler measurement.

\subsection{The WizCOLA N-body simulations}

In order to obtain an accurate covariance matrix for our BAO fits, we
require a much larger suite of N-body simulations than we used in the
previous sections.  We therefore analyzed the ``WizCOLA'' simulations
(\cite{Kodaetal2015}), which were produced in order to extend the
Comoving Lagrangian Acceleration (COLA) technique \citep{Tassev2013}
to lower-mass haloes to enable reconstruction of the baryon acoustic
peak in the WiggleZ Dark Energy Survey (\cite{Kazinetal2014}).  The
simulations are generated within a $600 \, h^{-1}$ Mpc box using the
``WMAP5'' cosmological model.  In particular, we used 1000 halo
catalogues output at $z=0$ in real-space, using halos with mass $M >
10^{13} \, h^{-1} M_\odot$ as above, which yields a catalogue with
number density $\sim 5 \times 10^{-4} \, h^3$ Mpc$^{-3}$,
characteristic of current and future large-scale structure surveys at
high redshift.

We applied the density-field reconstruction method to these simulation
boxes using standard methods described by,
e.g.\ \cite{Padmanabhanetal2012,Andersonetal2014,Kazinetal2014}.  The
process involves applying the displacements to both the data and a set
of random points, which we generated uniformly within each simulation
box with a number density 10 times greater than that of the data
itself.  The post-reconstruction correlation function is then measured
using the displaced data and random points.  When determining the
displacement field we assumed a linear bias factor $b \approx 1.4$,
fixed using the large-scale halo power spectrum.

\subsection{The correlation functions measured in different environments}

\begin{figure}
\begin{center}
\includegraphics[scale=0.4]{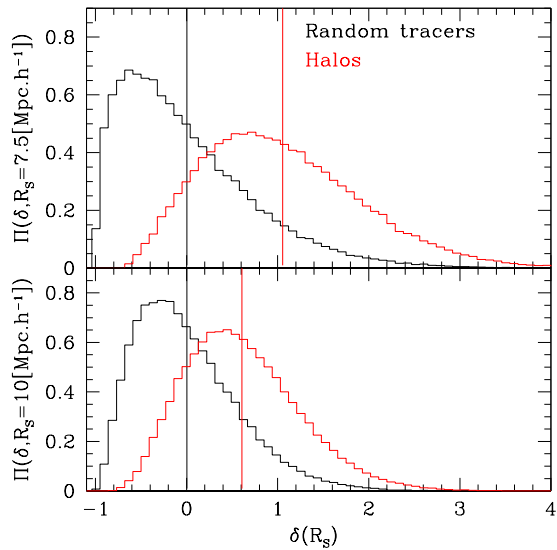}
\caption{The probability density function for the smoothed overdensity
  $\delta$ around halos and random tracers measured at $z=0$.  The top
  panel shows this PDF for $R_S = 7.5 \, h^{-1}$ Mpc, while the lower
  panel is for $R_S = 10 \, h^{-1}$ Mpc.}
\label{Fig7}
\end{center}
\end{figure}

As in section \ref{secenv}, we divided the simulation into different
density regimes by smoothing the density field traced by the halos.
Fig.(\ref{Fig7}) displays the probability density function
$\Pi(\delta,R_S)$ of the local overdensity $\delta$ around halos (red
histogram) and random tracers (black histogram) at $z=0$. The top
panel is for a smoothing scale $R_S = 7.5 \, h^{-1}$ Mpc, while the
lower panel shows the PDF for $R_S = 10 \, h^{-1}$ Mpc.  The vertical
lines show the mean of the two distributions; for random tracers the
mean is always zero, while for halos the mean of the distribution
increases for smaller smoothing scales due to clustering.

We split our sample into different environment bins using the local
density around halos.  For the purposes of this section we use 5
different environments.  We defined the bin divisions by measuring the
mean $\overline{\delta}$ and standard deviation $\sigma_\delta$ of the
density values for each halo.  In this case, environment $E_3$
corresponds to the average density and the central density values of
the other environments are set using
\begin{equation}
 \delta = \bar{\delta} + (i-3) \sigma_\delta.
\end{equation}
We also tested defining the bins based on the local density around
random tracers, which produced qualitatively similar results.

We measured auto- and cross-correlation functions between different
environments using the Landy-Szalay estimator
(\cite{LandySzalay1993}). The cross-correlation function between
environments $(E_i, E_j)$ is therefore
\begin{equation}
\xi_{ij} = \frac{DD_{ij}}{RR_{ij}} \frac{n_{Ri} n_{Rj}}{n_{Di} n_{Dj}}
- \frac{DR_{ij}}{RR_{ij}} \frac{n_{Ri}}{n_{Di}} -
\frac{DR_{ji}}{RR_{ij}} \frac{n_{Rj}}{n_{Dj}} + 1
\label{xi_ij}
\end{equation}
 where $DD_{ij}$ are the pair counts of halos
between environments $(i,j)$ $n_{Di}$ is the total number of halos in
the environment $i$, $RR_{ij}$ are the pair counts of random tracers,
$DR_{ij}$ the cross-pair counts, and $n_{Ri}$ the total number of
random tracers in environment $i$.

We determined the covariance matrix of the cross-correlation function
between two separation bins $(r_k, r_l)$ by averaging over the
ensemble of N-body realizations:
\begin{equation}
C_{k,l} = \langle \xi_{ij}(k) \, \xi_{ij}(l) \rangle - \langle
\xi_{ij}(k) \rangle \langle \xi_{ij}(l) \rangle
\label{Cov}
\end{equation}
The error in the measurement of $\xi_{ij}(r_k)$ is then $\sqrt{C_{k,k}}$.

\begin{figure}
\begin{center}
\includegraphics[scale=0.4]{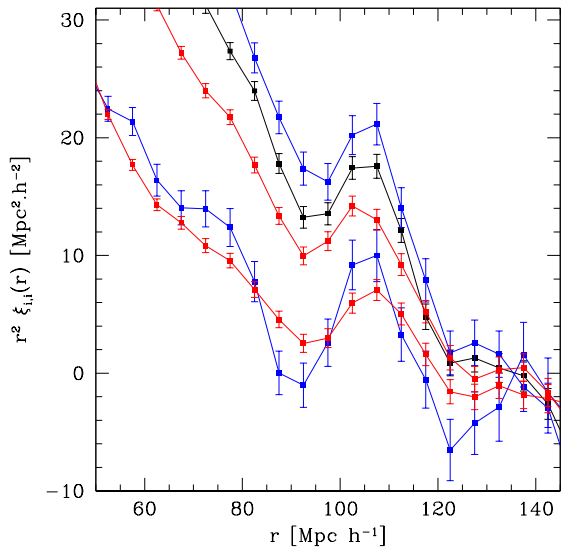}
\caption{The auto-correlation functions measured in the different
  environments, from bottom to top the measurements correspond to
  $\xi_{11}, \xi_{55}, \xi_{44}, \xi_{33}, \xi_{22}$. We use $R_S = 10
  \, h^{-1}$ Mpc for the smoothing scale.}
\label{Fig8}
\end{center}
\end{figure}

In Fig.(\ref{Fig8}) we plot the auto-correlation functions
$\xi_{ii}(r)$ measured in the 5 different environments after
reconstruction.  The black line (and dots) correspond to $E_3$, the
red lines (and dots) correspond to overdense environments $E_4,E_5$
while the blue lines (and dots) are for underdense environments
$E_1,E_2$.  In order to produce a clearer visualization for the
purposes of this plot, we stacked the simulation boxes in groups of 10,
such that the error in the measurements is then determined from 100
realizations of 10 stacked boxes.  Fig.(\ref{Fig8}) clearly displays a
hierarchy where the baryon acoustic peak is sharper for the underdense
regions and becomes more blurred for the high density regions.
Interestingly, the most under-dense environment produces a ``bell
shape'' around the acoustic peak.  The cross-correlation functions
display behaviour intermediate between the corresponding
auto-correlation functions.

\begin{figure}
\begin{center}
\includegraphics[scale=0.4]{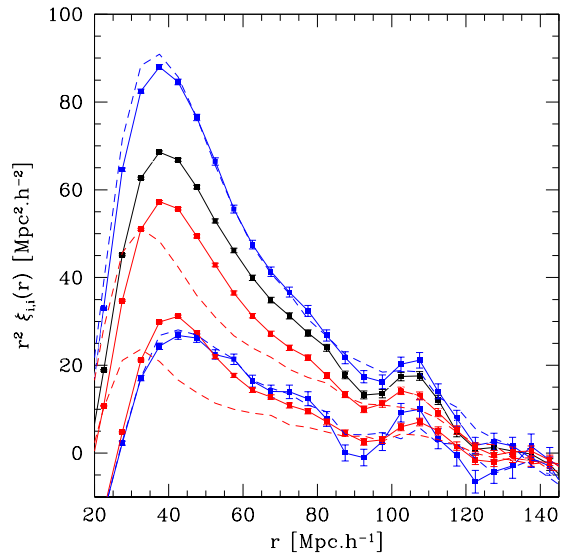}
\caption{Correlation functions measured in the different environments
  for $R_S = 10 \, h^{-1}$ Mpc before (dashed lines) and after
  reconstruction (solid lines and dots).  From bottom to top the
  measurements correspond to environments $E_1, E_5, E_4, E_3, E_2$.}
\label{Figbfaf}
\end{center}
\end{figure}

In order to further understand how the correlation functions
$\xi_{ij}$ are affected by the reconstruction method, we plot in
Fig.(\ref{Figbfaf}) the full correlation function before and after
reconstruction.  We see that reconstruction uniformly improves the
sharpness of the acoustic peak in all environments.

The increased sharpness of the post-reconstruction baryon acoustic
peak in underdense environments could potentially be used to improve
the accuracy of the standard ruler by upweighting those environments.
However, the fluctuating weight could also serve to increase the
covariance in our measurements.  We now explore the trade-off between
these effects.  We note that using density-dependent weights will
always result in a biased estimator for the underlying correlation
function.  However, given that the amplitude and shape of the
correlation function are marginalized over in standard methods of
fitting the preferred scale, this bias is not necessarily problematic
if the acoustic peak itself is not shifted.

\subsection{Total correlation function and new estimators}

The total correlation function $\xi$ can be expressed as a function of
the environmental correlation functions Eq.(\ref{xi_ij}).  We begin
with the Landy-Szalay estimator for $\xi$,
\begin{equation}
\xi = \frac{DD}{RR} \frac{n_R^2}{n_D^2} - 2 \frac{DR}{RR}
\frac{n_R^2}{n_{DR}^2} + 1
\label{LSstand}
\end{equation}
where $DD$, $RR$ and $DR$ are the total pair counts (halo-halo,
random-random tracer, halo-random tracer). The total number of halos
($n_D$), random tracers ($n_R$) and cross-product ($n_{DR}$) appearing
in this equation satisfy $n_D^2 = \sum_{ij} n_{Di} n_{Dj}$, $n_R^2 =
\sum_{ij} n_{Ri} n_{Rj}$ and $n_{DR}^2 = \sum_{ij} n_{Di} n_{Rj}$,
where the sum is over all environments (and is written in this manner
to allow us to generalize the relation as developed below). The
conservation of the total number of pairs can be expressed as
\begin{equation}
DD = \sum_{i,j} DD_{ij}
\label{conser}
\end{equation}
Substituting Eq.(\ref{conser}) into Eq.(\ref{LSstand}), we can express
$\xi$ as function of the $\xi_{ij}$ as
\begin{equation}
\xi = \frac{n_R^2}{n_D^2} \frac{1}{RR} \sum_{i,j} DD_{ij} - 2
\frac{DR}{RR} \frac{n_R^2}{n_{DR}^2} + 1
\label{xitot}
\end{equation}
where 
\begin{equation}
\begin{split}
&\sum_{i,j} DD_{ij} \equiv RR_{ij} \frac{n_{Di} n_{Dj}}{n_{Ri} n_{Rj}} \left(
  \xi_{ij} - 1 \right) -
  \\ & DR_{ij} \frac{n_{Di}}{n_{Ri}} - DR_{ji} \frac{n_{Dj}}{n_{Rj}} .
\end{split}
\end{equation}
Hence we have obtained an expression for the total correlation function
in terms of $\xi_{ij}$.

\subsubsection{Illustrative example}
\label{illu}

As a first illustration of the effect of assigning a different weight
to different environments, we re-write $\xi_{ij}$ in Eq.(\ref{xitot})
as $w_{ij} \, \xi_{ij}$, where $w_{ij} \equiv \sqrt{w_i w_j}$ is the
weight assigned to each cross-correlation function in terms of weights
$w_i$ defined for each environment $E_i$.  In the case where $w_i =
1$, we recover the original correlation function.  For the purposes of
this example we consider a simpler split into 3 environments.

\begin{figure}
\begin{center}
\includegraphics[scale=0.4]{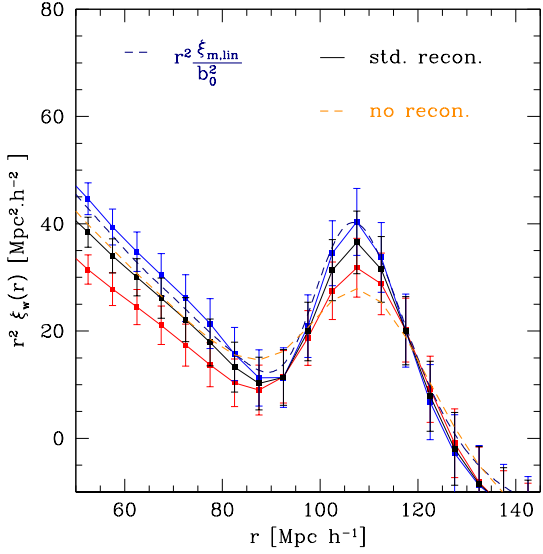}
\caption{The linear correlation function (blue dashed line) and
  measured correlation function at $z=0$ (orange dashed line). The
  standard reconstructed correlation function is shown by the black
  line, while the blue and red lines are particular cases of our new
  estimate for the reconstructed correlation functions weighting by
  environment.}
\label{xitotw1}
\end{center}
\end{figure}

In Fig.(\ref{xitotw1}), we show the linear matter correlation function
rescaled by the linear bias (blue dashed line) and the
non-linear matter power spectrum measured in the simulations (orange
dashed line). The black line (and dots) shows the reconstructed
correlation function using $R_S = 10 \, h^{-1}$ Mpc with no weighting.
For comparison we plot the weighted correlation function for two
special cases (blue and red), where we choose $w_1=2, w_2=2, w_3=0.5$
for the blue line (upweighting low-density environments) and $w_1=0,
w_2=0.5, w_3=2$ for the red line (upweighting high-density
environments).  We can see that weighting the underdense environment
($E_1$) more than the overdense environment ($E_3$) allows us to
recover the linear (sharpest) amplitude of the BAO peak.

\subsubsection{General linear weighting}

We now construct a general estimator for the weighted correlation
function, insensitive to the absolute value of the weights, as
\begin{equation}
\xi_w = \frac{\sum_{ij} \left( w_{ij} \, \xi_{ij} \, RR_{ij} \,
  \alpha_{ij} + \beta_{ij} \right)}{\sum_{ij} w_{ij} \, RR_{ij}}
\label{xitotw}
\end{equation}
where
\begin{equation}
\alpha_{ij} = \left( \frac{n_R}{n_D} \right)^2 \frac{n_{Di} \,
  n_{Dj}}{n_{Ri} \, n_{Rj}},
\end{equation}
and
\begin{equation}
\begin{split}
&\beta_{ij} = \left( \frac{n_R}{n_D} \right)^2 \bigg[ \Big( \frac{n_{Dj}}{n_{Rj}}- \frac{n_D^2}{n_{DR}} \Big) DR_{ij} + \\
&\Big( \frac{n_{Di}}{n_{Ri}} - \frac{n_D^2}{n_{DR}}\Big)DR_{ji} + \left( \frac{n_{Di} \, n_{Dj}}{n_{Ri} \, n_{Rj}} - \frac{n_D^2}{n_R^2} \right) RR_{ij} \bigg]  
\end{split}
\end{equation}
and we re-defined $n_D^2 = \sum_{ij} w_{ij} n_{Di} n_{Dj}$, $n_R^2 =
\sum_{ij} w_{ij} n_{Ri} n_{Rj}$ and $n_{DR}^2 = \sum_{ij} w_{ij}
n_{Di} n_{Rj}$.  This new estimator has the following desirable
properties: $\xi_w = \xi$ for $w_{ij} = {\rm constant}$, and $\xi_w =
\xi_{kl}$ when $w_{ij} = \delta_{ik}\delta_{jl}$.\\
 
To assign a value for each $w_{ij} = \sqrt{w_i w_j}$ in
Eq.(\ref{xitotw}) we used the following parametrization:

\begin{equation}
w_i \equiv 1 + x \left( \frac{i - \bar{i}}{i_{\rm max} - \bar{i}} \right)
\label{wieq}
\end{equation}
where $\bar{i}$ corresponds to the environment for which the local
density $\delta(R_S) = \bar{\delta}$, $i_{\rm max} = 2 \bar{i} -
i_{\rm min}$ is the highest density environment, and $x$ is a variable
that varies in the range $x = \left[-1,1\right]$ such that for $x=-1$,
the lowest-density environment is weighted with the highest amplitude
$w_{i_{\rm min}} = 2$ and the highest environment is weighted with
$w_{i_{\rm max}}=0$.

\begin{figure}
\begin{center}
\includegraphics[scale=0.4]{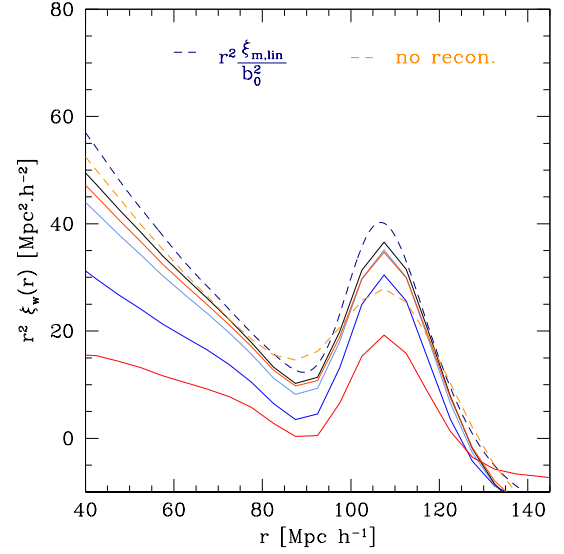}
\caption{The linear correlation function (blue dashed line) and
  measured unweighted correlation function at $z=0$ (orange dashed
  line). The solid lines show the weighted correlation function for
  $x=-1,-0.5,0,0.5,1,$ which correspond to the blue, dark blue,
  black, orange and red lines, respectively.  We do not show the
  errors in the measurement for clarity of the Figure.}
\label{xitotw2}
\end{center}
\end{figure}

In Fig.(\ref{xitotw2}) we can see the resulting weighted correlation
functions for various values of $x$, as well as the linear and
non-linear correlation functions.  We used 5 different environments
and $R_S = 10 \, h^{-1}$ Mpc.  We observe that the hierarchy of the
$\xi_w$ is not as simple as we show in Fig.(\ref{xitotw1}). With this
parametrization, the standard reconstruction (black line, $x=0$) is
closer to the linear correlation function than the measurement
obtained assigning higher weight to the underdense environments (blue
and light-blue lines) owing to the amplitude bias introduced by the
weighting.  However, the sharpness of the acoustic peak is nonetheless
increased by such a weighting scheme.  In the next section we consider
the implications for the recovered standard ruler scale.
 
\subsection{Baryon acoustic peak fitting}
\label{alphafitsec}

In order to determine the accuracy of the recovered distance scale, we 
fit a simple baryon acoustic peak model to the galaxy correlation
functions measured from each N-body realization, using the distribution
of measurements across the realizations as the covariance matrix.  Our
fiducial correlation function model is based on a Fourier-transform of a 
model linear power spectrum $P(k)$:
\begin{equation}
\xi_{\rm fid}(s) = \frac{1}{2\pi^2} \int dk \, k^2 \, P(k) \, \left[
\frac{\sin{(ks)}}{ks} \right] .
\end{equation}
We constructed the model power spectrum using a transfer function from
the Eisenstein \& Hu (1998) fitting formulae, assuming the fiducial
cosmological parameters of the WiZCOLA simulations.  We then fit for
the scale distortion parameter $\alpha$, marginalizing over a
normalization factor $b^2$:
\begin{equation}
\xi_{\rm mod}(s) = b^2 \, \xi_{\rm fid}(\alpha s)
\end{equation}
This approximate model is adequate for our purposes of exploring the
relative accuracy of determining $\alpha$; we verified that a series
of model extensions, such as using a model power spectrum from CAMB,
or marginalizing over an additive scale-free polynomial or a free
damping parameter, or using the mock mean correlation function as a
template, did not qualitatively change our conclusions.

We quantified the standard ruler performance by the standard deviation
of the best-fitting values of $\alpha$ across the realizations, which
we denote by $\sigma_\alpha$.  For this particular test we used the
1000 single-box COLA realizations and 12 environments, checking that
our conclusions were not sensitive to these choices.

In order to test any improvement brought by the new concepts developed
in the previous sections, we defined as a reference level the results
obtained by a standard reconstruction method using the Zel'dovitch
approximation, smoothing length $R_S = 10 \, h^{-1}$ Mpc, and no
weighting ($x=0$).  Similar assumptions are used in standard
implementations such as
(\citep{Padmanabhanetal2012,Andersonetal2014,Kazinetal2014}).  Using
the measurements of $\sigma_\alpha$, we computed the relative
difference $\Delta \sigma_\alpha$ between the reference model and the
results obtained (a) changing the smoothing length, (b) adding the
2LPT correction and (c) introducing weighting of environments as a
function of $x$.

\begin{figure}
\begin{center}
\includegraphics[scale=0.4]{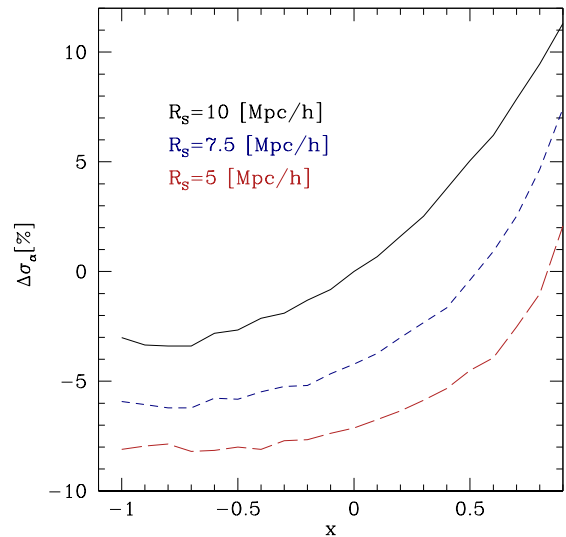}
\caption{The accuracy of the recovery of the standard ruler after
  density-field reconstruction, relative to a reference
  implementation, as a function of the parameter $x$ in
  Eq.(\ref{wieq}) which controls the relative weight assigned to
  underdense and overdense regions. The reference case is the
  ZA with $R_S = 10 \, h^{-1}$ Mpc and no weighting.}
\label{Fig9}
\end{center}
\end{figure}

In Fig.(\ref{Fig9}) we display our results for this analysis, in the
form of the relative difference $\Delta \sigma_\alpha$ as function of
the weighting parameter $x$ defined by Eq.(\ref{wieq}).  The solid
line uses the reconstruction method with the Zel'dovitch approximation
and $R_S = 10 \, h^{-1}$ Mpc, the blue short dashed line shows the
result for 2LPT and $R_S = 7.5 \, h^{-1}$ Mpc while the red dashed
line shows 2LPT and $R_S = 5 \, h^{-1}$ Mpc.

We find that the environmental-dependent weighting produces a small
but measurable improvement in $\sigma_\alpha$ ($\sim 3\%$), with best
performance usually produced in the range $-1 < x < -0.5$.  The
implementation of 2LPT for computing the displacement field also
produces a benefit as judged by $\sigma_\alpha$, such that the total
improvement is $\sim 8\%$.

\section{Discussion and conclusion}

We summarize the conclusions of our study as follows:

\begin{itemize}

\item The displacement of proto-halos can be predicted with high
  accuracy by the ZA using their peculiar velocities.  In this case,
  Eq.(\ref{Psi_Zv}) evaluated at $R_S=R_L$ provides a very good
  description of the displacement field with accuracy $\sigma_\Psi
  \sim 0.5 \, h^{-1}$ Mpc, independently of the halo masses.  The
  choice of the smoothing scale is important, since there is a unique
  scale for which the approximation gives the best agreement with the
  exact final positions of the halos.  The error in each component of
  $\mathbf{\Psi}$ follows a Gaussian distribution centered on zero
  with a variance sensitive to the smoothing scale and the environment
  of the halos.

\item When reconstruction of the displacement field is performed using
  the $z=0$ halo distribution in real space, we likewise find that the
  accuracy depends significantly on the smoothing scale and
  environment.  We extended the ZA to 2LPT and established a link
  between the error in predicting the displacement field and the local
  density around halos.

\item Based on this result, we showed that after applying
  reconstruction, the baryon acoustic peak is sharper in the
  correlation function of low-density environments.  We hence built a
  new estimator of the correlation function, constructed by weighting
  a set of auto- and cross-correlation functions measured between
  different environments.

\item Fitting a BAO model to the results, we found a small but
  measurable improvement of $\sim 8\%$ in determining the standard
  ruler scale through a combination of using 2LPT to find the
  displacement field and weighting the environmental correlation
  functions.  Further improvements may be possible using more
  sophisticated weighting schemes.

\end{itemize}

The fact that the reconstruction of the displacement field using the
halo density distribution at $z = 0$ is $\sim 5$ times less accurate
than the ZA applied to the initial velocities of proto-halos
(independently of the orders considered in the LPT approximation)
implies that the measurement of the error $\sigma(\psi_i -
\psi_{i}^{(1)})$ contains non-linear information which could
potentially be sensitive to different cosmological models, especially
in high-density environments.  The evolution in time of the
environmental correlation functions should carry this non-linear
information and can be use to test non-standard cosmologies.

For instance in Fig.(\ref{Figbfaf}), the correlation function before
reconstruction in high-density environments has a smeared BAO peak
compared to that of the lowest density environment.  The ratio of the
BAO peak widths in those two environments should be directly
proportional to the degree of late time non-linear interactions.  For
non-standard cosmologies, this ratio might be different (e.g., fewer
non-linear interactions for early dark energy models).  We will
investigate these topics further in future work.

\section{Acknowledgments}

We warmly thank Y. Rasera for providing easy access to the DEUS
simulations\citep{DEUSweb}. We are also grateful to Jun Koda for
generating the WizCOLA simulation suite used in this investigation.
Part of this research was conducted by the Australian Research Council
Centre of Excellence for All-sky Astrophysics (CAASTRO), through
project number CE110001020. We also acknowledge support from the DIM
ACAV of the Region Ile-de-France.

\end{document}